\documentclass[graphicx,times]{mn2e}
\usepackage{epsfig}
  
\newcommand{\gsim}{\mathrel{\hbox{\rlap{\lower.55ex \hbox {$\sim$}}
                   \kern-.3em \raise.4ex \hbox{$>$}}}}

\title[Variations in the sub-stellar IMF]{The dependence of the
sub-stellar IMF on the initial conditions for star formation}

\author[Delgado-Donate, Clarke \&
Bate]{E. J. Delgado-Donate$^1$\thanks{E-mail: edelgado@ast.cam.ac.uk},
C. J. Clarke$^1$, M. R. Bate$^2$\\ $^1$Institute of Astronomy,
University of Cambridge, Madingley Road, Cambridge, CB3 0HA\\ $^2$
School of Physics, University of Exeter, Stocker Road, Exeter EX4 4QL}

\begin{document}

\date{}

\pagerange{\pageref{firstpage}--\pageref{lastpage}} \pubyear{2003}

\maketitle

\label{firstpage}

\begin{abstract}
We have undertaken a series of hydrodynamical simulations of multiple
star formation in small turbulent molecular clouds. Our goal is to
determine the sensitivity of the properties of the resulting stars and
brown dwarfs to variations in the initial conditions imposed. In this
paper we report on the results obtained by applying two different
initial turbulent velocity fields. The slope of the turbulent
power-law spectrum $\alpha$ is set to $-3$ in half of the calculations
and to $-5$ in the other half. We find that, whereas the stellar mass
function seems to only be weakly dependent on the value of $\alpha$,
the sub-stellar mass function turns out to be more sensitive to the
initial slope of the velocity field. We argue that, since the role of
turbulence is to create substructure from which gravitational
instabilities may grow, variations in other initial conditions that
also determine the fragmentation process are likely to affect the
shape of the sub-stellar mass function as well. The absence of many
planetary mass {\it free-floaters} in our simulations, especially in
the mass range $1-10$ M$_{\rm J}$, suggests that, if these objects are
abundant, they are likely to form by similar mechanisms to those
thought to operate in quiescent accretion discs, instead of via
instabilities in gravitationally unstable discs. We also show that the
distribution of orbital parameters of the multiple systems formed in
our simulations are not very sensitive to the initial conditions
imposed. Finally, we find that multiple and single stars share
comparable kinematical properties, both populations being able to
attain velocities in the range $1-10$ km s$^{-1}$. From these values
we draw the conclusion that only low-mass star-forming regions such as
Taurus-Auriga or Ophiuchus, where the escape speed is low, might have
suffered some depletion of its single and binary stellar population.

\end{abstract}


\section{Introduction}

It is of central importance in astrophysics to understand how stars
form and which mechanisms shape their properties. In particular, the
determination of the origin and functional form of the initial mass
function of stars and brown dwarfs (IMF) has become the holy grail of
star formation studies. The first statistically accurate derivation of
the IMF for field stars (Salpeter 1955) yielded a power-law functional
form d$N$/d$log$M $\propto$ M$^\gamma$ with slope $\gamma = -1.35$, in
the range $0.4-10$ M$_\odot$. Subsequent measurements of the IMF for
field stars, which explored a wider range of masses, have confirmed
the early results of Salpeter: e.g. Miller \& Scalo (1979)
approximated the IMF by a half-lognormal distribution, for masses
between 0.1 and $\approx 30$ M$_\odot$, the slope above 1 M$_\odot$
stars being very similar to Salpeter's. Kroupa (2001) defined an {\it
average or Galactic-field} IMF which also had a Salpeter slope above
0.5 M$_\odot$, but could be better fitted by a slope $\gamma = -0.3$
between 0.08 and 0.5 M$_\odot$ and $\gamma = +0.7$ in the sub-stellar
regime. Finally, Chabrier (2003) found that, as a general feature, the
IMF is well described by a power-law form for M~$\geq 1$~M$_\odot$ and
a lognormal form below, except possibly for early star formation
conditions. There is also evidence (see reviews by Kroupa 2002 and
Chabrier 2003) that, within the empirical errors, the IMF of clusters,
both open and globular, and young associations such as Taurus-Auriga
and Ophiuchus, also resembles closely that of field stars, with
perhaps some possible variations at the low-mass end. The lack of
clear evidence for IMF variations has raised the possibility that the
IMF for stars, at least for the disc populations at z $\sim 0$, might
be indeed universal.

The IMF at the sub-stellar regime is by no means so well
constrained. Brown dwarfs were not discovered until 1995 (Nakajima et
al. 1995; Rebolo, Zapatero-Osorio \& Mart\'{\i}n 1995), and since then
a lot of observational effort has been devoted to pinning down the
form of the IMF below the hydrogen burning limit (Delfosse et
al. 1999; Burgasser et al. 2000; Kirkpatrick et al. 2000; Leggett et
al. 2000). From these studies, Chabrier (2002, 2003) concluded that
the number density of Galactic disc brown dwarfs is comparable to that
of stars, and that the functional form of the IMF in the sub-stellar
regime can be characterised, within the uncertainties, by a lognormal
distribution. Recently, however, some results seem to indicate that
the sub-stellar IMF might indeed be more sensitive to {\it
environmental conditions} than the stellar IMF. On the one hand,
Jameson et al. (2002) find an IMF slope of $\gamma \approx 0.4$ in the
sub-stellar regime of the Pleiades for the mass range 0.02-0.075
M$_\odot$, Muench et al. (2002) show that the Orion-Trapezium cluster
has a brown dwarf fraction of $\approx 30\%$ in the same mass range
and Barrado y Navascu\'es et al. (2002) derive a sub-stellar IMF slope
of $\gamma \approx 0.6$ for the $\alpha$ Persei cluster; i.e. the
Pleiades, Trapezium and $\alpha$ Persei young clusters seem to contain
a number of brown dwarfs comparable to that of stars. On the other
hand, Luhman et al. (2000) and Brice\~no et al. (2002) find that brown
dwarfs are $2 \times$ less frequent in Taurus-Auriga than in Orion,
result very similar to that found in the IC-348 cluster by Luhman et
al. (2003) and Preibisch, Stanke \& Zinnecker (2003).

Theoretically, the investigation of the origin of the IMF is by no
means straightforward. First of all it is necessary to model the
formation of a large number of stars so that the corresponding
theoretical IMF can be meaningfully compared with observations. One of
the first such attempts was made by Bonnell and collaborators (Bonnell
et al. 1997; 2001), who modelled clouds of gas with stellar {\it
seeds} randomly placed throughout that yielded a functional form for
the IMF in close resemblance to the observed stellar IMF. Bonnell et
al. (1997, 2001), as well as Klessen, Burkert \& Bate (1998) and
Klessen \& Burkert (2000, 2001), identified two physical processes
that contributed to shape the IMF, i.e. dynamical interactions between
protostars and competitive accretion. Gravitationally unstable groups
of stellar {\it seeds} (or dense cores in the case of Klessen \&
Burkert simulations) would start off with very small masses and
subsequently would grow in mass by accretion, which was dubbed
competitive because all {\it seeds}/cores attempt to feed from the
same gas reservoir. This approach was explored further by
Delgado-Donate, Clarke \& Bate (2003), who performed a large number of
calculations of small-$N$ clouds, set up in a similar manner as
Bonnell et al. (1997). They concluded that the IMF at the stellar
regime is likely to follow the parent core mass function down to the
minimum core mass, just to flatten below where internal processes
within each core would be dominant.

A more direct approach has recently been conducted by Bate, Bonnell \&
Bromm (2003; henceforth BBB), who performed a calculation of the
fragmentation and collapse of a 50 M$_\odot$ gas cloud that fully
resolved the opacity limit for fragmentation. This simulation, in
which a supersonic turbulent velocity field is initially imposed on
the gas, results in the formation of $\approx 50$ stars and brown
dwarfs, and demonstrated that the resulting IMF is compatible with
current observational measurements. This sort of calculations,
however, are very demanding computationally, and a wide range of
initial conditions cannot easily be explored. A different approach is
needed if the dependence of the functional form of the IMF on
different initial conditions (i.e. different star formation
environments) is to be studied. In this paper we have taken an
alternative approach which is a natural complement to the BBB
simulation. We find that by modelling an ensemble of isolated cores of
mass 5~M$_\odot$, we can improve the number of stars formed per CPU
hour by a factor 7 compared with the BBB calculation. This economy
stems from the fact that by focusing on individual dense cores, we
dispense with the computational expense of following the diffuse gas
in the BBB simulation. We have performed 10 different calculations in
which a turbulent velocity field is initially imposed on the gas. This
turbulent field is characterised by a power-law velocity spectrum with
slope $\alpha$. We have used a different $\alpha$ ($-3$ and $-5$) for
each half of the set of calculations. This way, we expect to asses the
sensitivity of the properties of the resulting stars and brown dwarfs
to the value of the slope of the initial turbulent spectrum in
particular, and to variations in the initial conditions for star
formation in general.

We stress that our simulations share the property of the BBB
simulation that they resolve the opacity limit for fragmentation (Low
\& Lynden-Bell 1976; Rees 1976) and that, assuming that fragmentation
does not occur at densities greater than those at which the gas
becomes opaque to infrared radiation, these calculations are able to
model the formation of all the stars and brown dwarfs that, under the
initial conditions imposed, can be produced. Our spatial resolution
limit for binaries allows us to study a wide range of separations, and
the particle numbers we employ allow us to model accretion discs
around the protostars which are as long lived as those modelled by
BBB.

Our study has four major findings:
\begin{itemize}
\item The slope of the sub-stellar IMF is sensitive to the initial
conditions imposed on the parent cloud. The $\alpha = -3$ case
produces a larger fraction of brown dwarfs than the $\alpha = -5$
case. The stellar IMF does not show such statistically significant
variations.
\item Few objects with masses below $\approx 0.01$ M$_\odot$ are
formed, despite the minimum fragment mass being $10 \times$
smaller. This is due to the high accretion rates typical of very young
objects formed out of direct collapse or disc fragmentation. Thus, if
a large population of free-floating planetary mass objects exists, two
explanations may be suggested: first, the standard values for the
opacity limit for fragmentation may need to be revised; and/or second,
planetary-mass free-floaters may be formed in quiescent discs at later
times when the disc mass is much smaller than that of the central
object.
\item The pattern of mass acquisition among single and multiple stars
is shown not to depend sensitively on the slope of the initial
turbulent power spectrum. The distribution of the orbital parameters
of multiples systems is similarly weakly dependent on initial
conditions.
\item Single and binary stars attain comparable velocities, between
$1-10$ km~s$^{-1}$. Higher-order multiples have lower velocity
dispersions. The $\alpha = -3$ case is more prolific in high-speed
escapers. Low-mass, loose star-forming regions such as Taurus or
Ophiuchus might have an overabundance of $N > 2$ multiples, as
lower-$N$ systems may easily escape the potential wells of these
associations.
\end{itemize}

The structure of this paper is as follows. In Section 2 the
computational method and initial conditions applied to our models are
described. Section 3 presents a description of the cloud fragmentation
process. The results on the IMF are given in Section 4. In Section 5
we discuss the dependence of the properties of multiple stars on the
initial conditions imposed. Our conclusions are given in Section 6.

\section{Computational method}

The calculations presented here were performed using a 3D hybrid SPH
$N$-body code, designed to follow the dynamics of the stellar system
resulting from the fragmentation and collapse of a cloud of cold gas
(Bate, Bonnell \& Price 1995). The SPH code is based on a version
originally developed by Benz (Benz 1990; Benz et al. 1990). The
smoothing lengths of particles are variable in space and time, subject
to the constraint that the number of neighbours for each particle
remains approximately constant at $N_{\rm neigh} = 50$. The SPH
equations are solved using a $2^{\rm nd}$-order Runge-Kutta-Fehlberg
integrator with individual timesteps for each particle (Bate, Bonnell
\& Price 1995). Gravitational forces and nearest neighbours are
calculated using a binary tree. We use the standard form of artificial
viscosity (Monaghan \& Gingold 1983) with strength parameters
$\alpha_{\rm v} = 1$ and $\beta_{\rm v} = 2$.

\subsection{Equation of state}

When the gravitational collapse of a molecular cloud core begins, the
density is still low enough to preserve an approximate balance between
compressional heating induced by the collapse and cooling by molecular
line emission (e.g. Larson 1969). During this stage, the gas
temperature $T$ can be considered to remain constant. Once the density
$\rho$ reaches a critical value $\rho_{\rm c}$ of $\approx 10^{-13}
{\rm g~cm}^{-3}$, the gas becomes optically thick to infrared
radiation and half the energy released by gravitational collapse is
retained as thermal energy: the gas is essentially adiabatic. This
critical density defines the so-called opacity limit for
fragmentation: for an adiabatic gas, with a ratio of specific heats
$\eta = 7/5$ (appropriate for diatomic gas), the product
($T^{3}\rho^{-1}$)$^{1/2}$ increases with density and, therefore, a
given Jeans-unstable\footnote{The Jeans mass for a gas of density
$\rho$ and temperature $T$ is given by ($\frac {5 R_{\rm g} T} {2 G
\mu}$)$^{3/2}$ ($\frac {4} {3}$ $\pi \rho$)$^{-1/2}$, where $R_{\rm
g}$ is the universal gas constant, $G$ the gravitational constant and
$\mu$ the mean molecular weight} collapsing clump quickly becomes
Jeans-stable, turning into a pressure-supported object.

This pressure-supported object initially contains a few Jupiter-masses
(M$_{\rm J}$) and has a radius of $\approx 5$ AU (Larson
1969). Although these objects later undergo another phase of collapse
due to the dissociation of molecular hydrogen (Larson 1969), it is
thought that they are unable to sub-fragment (Boss 1989; Bate
1998). Thus, the opacity limit sets a minimum fragment mass of a few
M$_{\rm J}$ (Low \& Lynden-Bell 1976; Boss 1988; BBB).

To model the opacity limit for fragmentation without performing full
radiative transfer, we use an equation of state given by $p$ = $K$
$\rho^{\eta}$, where $p$ is the pressure and $K$ is a measure of the
entropy of the gas. The value of $\eta$ changes with density as:
\begin{equation}
\eta = \cases{\begin{array}{rl}
1, & \rho \leq 10^{-13}~ {\rm g~cm}^{-3}, \cr
7/5, & \rho > 10^{-13}~ {\rm g~cm}^{-3}. \cr
\end{array}}
\end{equation}
The gas is assumed to consist of pure molecular hydrogen ($\mu$ = 2),
and the value of $K$ is such that when the gas is isothermal, $K$ =
$c_{\rm s}^2$, with $c_{\rm s} = 1.85 \times 10^4 {\rm cm~ s}^{-1}$ at
$T = 10$~K. The pressure is continuous when the value of $\eta$
changes.

This equation of state matches closely the relationship between
temperature and density obtained by full frequency-dependent radiative
transfer models of the spherically-symmetric collapse of molecular
cloud cores (Masunaga, Miyama \& Inutsuka 1998; Masunaga \& Inutsuka
2000). It should model collapsing regions well but where departure from
spherical symmetry becomes important (e.g. protostellar discs) it may
not model the thermodynamics particularly accurately.

\subsection{Sink particles}

Once the opacity limit for fragmentation is reached and a
pressure-supported fragment is formed, the integration of the SPH
equations within that object will merely slow down the calculation,
thus preventing us from studying the dynamical evolution of the rest
of the cloud. Assuming that fragmentation above the opacity limit is
irrelevant, the evolution within a collapsed fragment can be ignored
and the object replaced by a sink particle (Bate, Bonnell \& Price
1995) of the same mass and momentum. This sink particle is inserted
when the central density of the fragment exceeds $\rho_{\rm s} =
10^{-10} {\rm g~cm}^{-3}$, well above the critical density $\rho_{\rm
c}$. Sink particles are point masses with an accretion radius, so that
any gas particle that falls into it and is bound to the point mass is
accreted. Sink particles interact with the gas only via gravity and
accretion. The typical initial mass of a sink particle is $\sim$ a few
M$_{\rm J}$, as a result of the critical density used for the opacity
limit for fragmentation. In the present calculations, the sink's accretion
radius $R_{\rm sink}$ is constant and equal to 5 AU. Therefore, discs
around sink particles will be resolved only if their radii $\gsim 10$
AU. 

The gravitational acceleration between two sink particles is Newtonian
for $r \geq 4$ AU, but is smoothed within this radius using spline
softening (Benz 1990). The maximum acceleration occurs at $r \sim 1$
AU; therefore, this is the minimum binary separation that can be
resolved. Sink particles are not permitted to merge.

It is worth noting that although the insertion of sink particles, as
any other numerical switch, introduces some error in the calculations
(a short-lived structure could have been wrongly identified as a
protostellar object), the density at which pressure-supported objects
are replaced by sinks is 3 orders of magnitude above the critical
density $\rho_{\rm c}$. By then, the collapsed fragment is
self-gravitating, centrally-condensed and roughly spherical, and has
had some time to interact in that form with its environment, thus
being potentially exposed to coalescence or disruption. In practice,
most pressure-supported objects formed in these calculations would
have collapsed to stellar densities were it not for the insertion of a
sink particle.

\begin{figure*}
\begin{minipage}{15cm}
%\caption{Figure too large for astro-ph version. It can be provided on request
%to first author.}
\caption{Snapshots of the column density structure of the cloud. On the
left for the $\alpha3$ calculations, on the right for the $\alpha5$
ones. The upper panels correspond to approximately 10000 yr after the
simulations begin. The bottom panels show the state of the cloud some
time after the first free fall time, once star-formation has
commenced. Column density is plotted in linear scale in the upper
panels (where the physical scale is $\approx 10^4$ AU), and in
logarithmic scale in the bottom panels (here the scale length is
$\approx 10^3$ AU). Note: higher resolution figure available on
request to first author.}
\begin{center}  
\includegraphics[height=15cm]{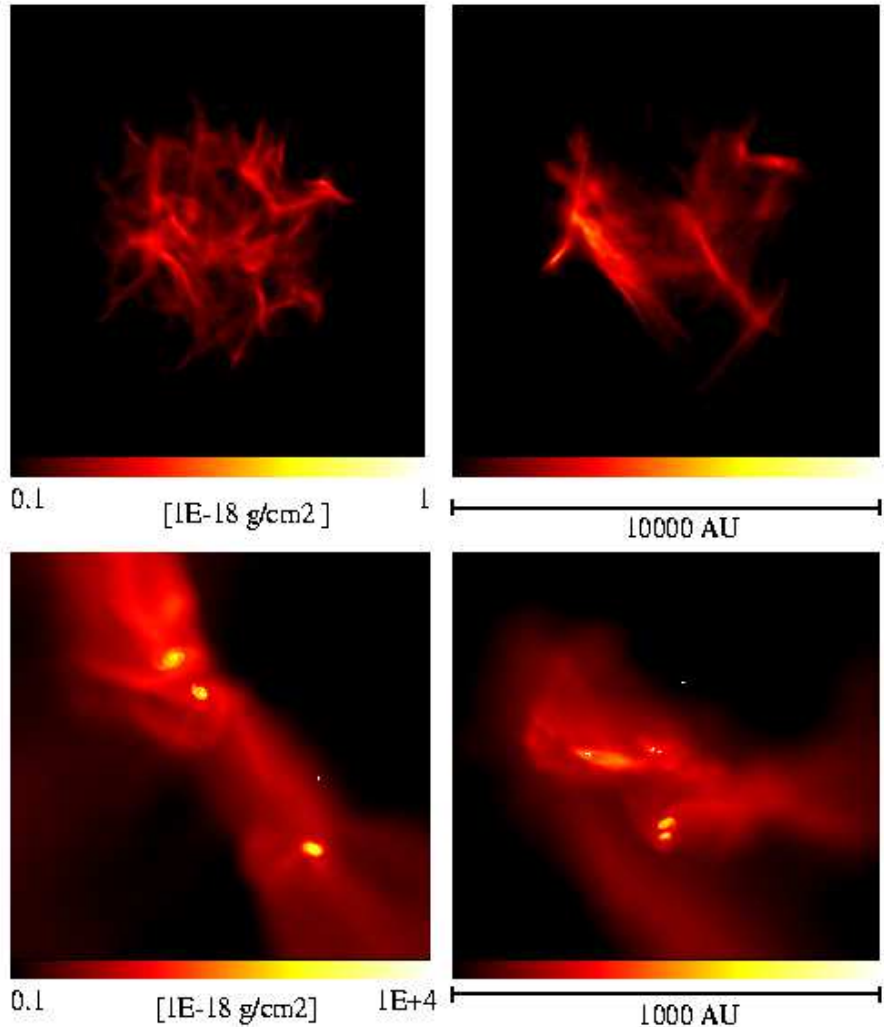}
\end{center}
\end{minipage}
\end{figure*}

\subsection{Initial Conditions}

We have performed 10 different calculations of the fragmentation of a
small-scale, turbulent molecular cloud, each of them under almost
exactly the same initial conditions. Each cloud core is spherical, has
a mass of 5 M$_\odot$, a radius of $\approx 10^4$ AU, and an initial
uniform density of $10^{-18} {\rm g~cm}^{-3}$. At the initial
temperature of 10 K, the mean thermal Jeans mass is 0.5 M$_\odot$,
i.e. the Jeans number of the cloud is 10. The global free-fall time of
the cloud $t_{\rm ff}$ is $\approx 10^{5}$ yr.

We have imposed an initial supersonic turbulent velocity field on the
gas, in the same manner as Ostriker, Stone \& Gammie (2001) and
BBB. We generate a divergence-free random Gaussian velocity field with
a power spectrum $P(k) \propto k^{\alpha}$, where $k$ is the
wavenumber and $\alpha$ is the power index, {\it which we have set to
$-3$ in half of the simulations and to $-5$ in the other half}. These
values of $\alpha$ bracket the empirical uncertainty in Larson's
velocity dispersion-size relation $\sigma \propto \lambda^{\zeta}$
(Larson 1981). Typical values for $\zeta$ range between $0.35$ and
$0.75$ (Larson 1981; V\'azquez-Semadeni et al 1997 and references
therein), the most quoted value being $0.4$. The exponent $\zeta$ and
$\alpha$ are not linearly related. Following Myers \& Gammie (1999),
we find that $\alpha = -3$ corresponds to $\zeta = 0.2$ whereas
$\alpha = -5$ corresponds to $\zeta = 0.8$. Larson's mass-size
relation M~$\propto \lambda^2$ is also satisfied by our models
(5~M$_\odot$ in 0.08~pc diameter clouds), although on the high-side of
the relation's scatter. Some of the Lynds clouds in Larson's original
paper, for example, have similar parameters to those used in these
models.

Each of the 5 calculations with the same $\alpha$ differ in the values
of the random numbers used to generate the velocity field. In turn,
for every set of random numbers used to generate a given velocity
field, calculations with both values of $\alpha$ have been
performed. The velocity field is normalised so that initially it is in
equipartition with the gravitational potential energy of the cloud,
i.e. initially the cloud is supported by the turbulent motions. The
initial root mean square (rms) Mach number of the flow is 3.75. This
value of the rms Mach number is in agreement with Larson's velocity
dispersion-size relation, albeit on the high side of the scatter. It
must be noted, however, that the turbulent velocity field imposed
initially in these calculations is mostly aimed at providing the seeds
for the growth of perturbations with a substantial amount of vorticity
(as opposed to the models by Delgado-Donate, Clarke \& Bate 2003), and
can be thought of as a remnant of the previous turbulent fragmentation
process (Padoan \& Nordlund 2002) that formed the cloud. Consequently,
it is not replenished further but permitted to decay freely. Thus, the
Mach number will quickly drop to (and remain for the rest of the cloud
evolution at) values closer to those observed in molecular cloud
cores.

The initial net angular momentum of the cloud is very small. If locked
in global solid body rotation, it would account for a value of $\beta$
(the ratio of rotational to gravitational potential energy in the
cloud) as low as 10$^{-3}$. Yet locally, the angular momentum can be
very high, as contained in the vorticity associated with the modes of
a divergence-free turbulent field. In principle, these initial
conditions do not favour the formation of wide binary stars.

Each cloud is bound by a small external pressure in order to prevent
the escape of SPH particles during the first free fall time of the
cloud evolution, when the thermal energy contained in the gas is not
negligible relative to the cloud gravitational energy. These
simulations do not include magnetic fields, as we have tried to
isolate a particular hydrodynamical fragmentation problem to
characterise the properties of the resulting stellar systems. We have
not included in our models any mechanical or radiative feedback
mechanism. This may be an appropriate choice, since the maximum
stellar mass in these simulations does not exceed 1 M$_\odot$, whereas
the most powerful winds and photoionisation fronts in star-forming
regions are produced by much more massive stars.

\subsection{Resolution}

The local Jeans mass must be resolved throughout the calculation (Bate
\& Burkert 1997; Truelove et al. 1997; Whitworth 1998), otherwise some
of the fragmentation might be artificially suppressed. The minimum
Jeans mass M$_{\rm res}$ occurs at the maximum density during the
isothermal collapse phase $\rho_{\rm c}$, and is $\approx 0.0015$
M$_\odot$ (1.5 M$_{\rm J}$). When modelling self-gravitating gas with
SPH, a Jeans mass must contain at least $2 \times N_{\rm neigh} = 100$
SPH particles (Bate \& Burkert 1997). Thus, following equation (6) of
Bate \& Burkert (1997), we need to use $3.5 \times 10^5$ particles to
model a 5 M$_\odot$ cloud core. In practice, any collapsing gas clump
in these simulations contains more SPH particles than M$_{\rm res}$,
since a clump with a mass of M$_{\rm res}$, when compressed, would
heat up and subsequently would exceed the Jeans mass: truly
pre-stellar clumps must be more massive than M$_{\rm res}$ to keep
collapsing. 

Angular momentum transport in accretion discs is also affected by the
number of particles used in the simulations. SPH artificial viscosity
forces (in particular the $\alpha_{\rm v}$ term) give rise to
unrealistically high values of the effective viscosity in accretion
discs (Artymowicz \& Lubow 1994). The consequence is that, unless many
particles (of the order of $10^5$) are present in the disc at any
given time -- which is not the case -- the disc dissipates too
quickly. This may have an effect on the dynamical interactions between
stars and/or brown dwarfs in our models: e.g. a given star-star
encounter may well have different outcomes depending on whether it is
mediated by discs or not, since the amount of dissipation of the
stars' kinetic energy depends on the amount of disc material they
interact with. In addition, since we can only resolve discs which are
larger than ~10 AU, encounters with impact parameters smaller than 10
AU will not be modelled completely accurately even if they involve
stars that still harbour large discs.

Each of the calculations that are discussed in this paper required
$\approx 4000$ CPU hours on the SGI Origin 3800 Computer of the United
Kingdom Astrophysical Fluids Facility (UKAFF).

\section{The evolution of the cloud}

The hydrodynamical evolution of the cloud produces shocks which
decrease the turbulent kinetic energy that initially supported the
cloud. In parts of the cloud, gravity begins to dominate and dense
self-gravitating cores form and collapse. These dense cores are the
sites where the formation of stars and brown dwarfs occurs. The
turbulence decays on the dynamical timescale of the cloud core (as
found by MacLow et al. 1998; Stone, Ostriker \& Gammie 1998; and BBB,
among others), and star formation begins just after 1 to 1.5 global
free-fall times $t_{\rm ff}$. The calculations are stopped when 60\%
of the gas has been accreted (i.e. at 60\% star-formation
efficiency). In terms of $t_{\rm ff}$ this means that, for the
$\alpha$ = $-3$ and $\alpha$ = $-5$ calculations (henceforth $\alpha3$
and $\alpha5$) , we follow the evolution of the cloud for $\approx 4
t_{\rm ff}$ and $5.5 t_{\rm ff}$, respectively (i.e. an average of
$\approx 0.5$ Myr). Altogether, at 60\% efficiency, the calculations
produce 145 stars and brown dwarfs: 85 in the 5 $\alpha3$ runs, and 60
in the other 5 $\alpha5$ runs.

The physical scale of the structures formed during the first $t_{\rm
ff}$ is markedly different for the two sets of initial
conditions. Figure~1 shows four snapshots of the column density
structure in the clouds, the upper panels corresponding to $\approx
0.1 t_{\rm ff}$. The $\alpha3$ case is depicted on the left, and the
$\alpha5$ on the right. The formation of filamentary or sheet-like
structures is common to both initial conditions. But clearly the
$\alpha5$ case, in which more kinetic power is stored in the long
wavelength modes ($\lambda > 1000$ AU) of the turbulent velocity field
(relative to the $\alpha3$ case), shows larger coherent structures,
and results in a more pronounced expansion of the initial gas spatial
distribution. Once a large fraction of the initial turbulent kinetic
energy is dissipated via shocks, dense pockets of gas begin to
form. The $\alpha5$ case produces, on average, more dense cores per
simulation: from 2 to 3; and these cores are more widely separated
than in the $\alpha3$ case, in which never more than 2 dense cores are
formed. The average number of protostars formed in each core varies by
a factor of $\approx 2$ depending on the initial conditions used
($\approx 4$ in the $\alpha5$ simulations and $\approx 7$ in the
$\alpha3$ ones). The formation of the first pressure-supported object
also takes longer, on average, for the $\alpha5$ case. Once the
formation of several dense cores has taken place, the subsequent
evolution of the cloud and the star formation process inside each core
is qualitatively independent of the value of $\alpha$ initially
imposed.

All the objects formed in these calculations start off with a mass
close to the opacity limit for fragmentation. Subsequently, they grow
in mass by accretion. Initially, a pressure-supported object forms
within each dense clump, first in isolation but soon surrounded by an
accretion disc. Initially the mass of the disc is comparable to, and
often greater than the mass of the central object. Thus, the disc is
prone to the appearance of gravitational instabilities which, in most
cases, result in the fragmentation of the disc into one or more
protostellar objects (Bonnell 1994; Bonnell \& Bate 1994; Burkert,
Bate \& Bodenheimer 1997; Whitworth et al. 1995). The formation of
this first star generally occurs in the lowest of the local potential
minima. Surrounding condensations with slightly lower gas densities
form additional stars (e.g. in the filaments whose intersection
generated the first dense clump). Both the stars and the residual gas
are attracted by their mutual gravitational forces and fall towards
each other (see Figure~1, bottom panels). The interactions between the
gas and the protostars dissipate some of the kinetic energy of the
latter (Bonnell et al. 1997), allowing the stellar objects to rapidly
come close to the initial star and its disc-born companions, to form a
high-density sub-cluster containing from 2 up to 8 stars: a small-$N$
cluster. This process repeats itself in other parts of the
cloud. Subsequently, sub-clusters are attracted to each other and
merge to form the final mini-cluster. Thus, the star formation process
is hierarchical in nature, as has been vividly illustrated (for a 1000
M$_\odot$ cloud) by Bonnell, Bate \& Vine (2003).

\begin{figure}
\begin{center}
\caption{Number of stars $+$ brown dwarfs formed in each
calculation. Squares refer to the $\alpha3$ group, asterisks to the
$\alpha5$ simulations. The solid line represents the mean number of
objects (for the combined dataset) and the dashed line the standard
deviation of the mean. The initial Jeans number of the clouds is 10.}
\centerline{\epsfig{file=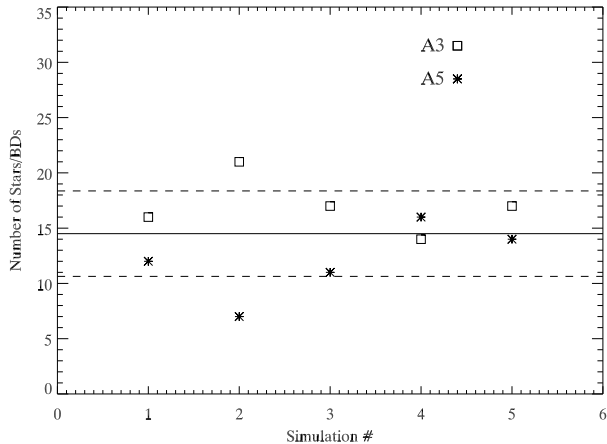,height=6.5cm}}
\end{center}
\end{figure}

Star formation is not only localised in space (the sub-clumps) but also
in time, i.e. proceeds in bursts (BBB): after one burst occurs, most
of the dense gas in the vicinity is exhausted and some time is needed
for more gas to be accreted from regions further away. The time
sequence of star-forming bursts depends on the local distribution of
dense gas and on the number and boundedness of the objects populating
the small-$N$ cluster. In most of the simulations the last burst is
triggered by the dynamical interactions induced by the last {\it
merger}, after which the cloud becomes stellar-dominated instead of
gas dominated. We have taken special care that all simulations were
run until at least 1 $t_{\rm ff}$ after the last star formation
event. Thus, we expect to have computed the formation of all the stars
and brown dwarfs that could ever have been formed in these clouds,
under the initial conditions imposed.

Overall, the $\alpha5$ simulations form less stars and brown dwarfs
than the $\alpha3$ runs (60 against 85). Stars and, particularly brown
dwarfs, form with a higher incidence if small-scale structure is
generated during the early evolution of the cloud. In this case
multiple fragmentation within a dense core occurs more
efficiently. The $\alpha5$ simulations are characterised by the
absence of small-scale structure initially, since there is
comparatively much less energy stored in small wavelengths ($\lambda <
1000$ AU) initially than in the $\alpha3$ case. In addition, it is the
long wavelength modes that contribute most to prevent the global
collapse of the cloud, therefore keeping dense cores away from each
other much longer: thus, star formation induced by mergers between
small-$N$ clusters occurs also more efficiently in the $\alpha3$
simulations than in the $\alpha5$ ones.

The process of fragmentation and collapse out of a turbulent molecular
cloud is largely stochastic. Hence, the number of stars and
brown dwarfs formed in each individual calculation varies. However, in
Figure~2 we show how this number seems to fluctuate around a mean
value, which coincides approximately (within a factor of 2) with the
initial number of Jeans masses contained in each cloud, i.e. 10. The
$\alpha5$ simulations form an average of $12 \pm 3$ stellar objects,
for $17 \pm 3$ in the $\alpha3$ group (the error being the standard
deviation of the mean). This result seems to suggest that in effect
the number of Jeans masses $N_{\rm J}$ may determine, at any rate, the
number of objects formed in a collapsing cloud: in principle there is
no {\it a priori} reason why some of the calculations could not have
produced, say, just one star or 50. It remains unclear however, how
far-reaching this relation between $N_{\rm J}$ and the number of
objects formed is (BBB also form $\approx 50$ objects in an initially
50 $N_{\rm J}$ cloud), and if it is linear or not.

\section{Variations in the sub-stellar mass function}

\begin{figure*}
\begin{center}
\caption{Mass functions. On the upper left, we show the mass function
derived from the $\alpha3$ calculations; on the upper right, the same
for the $\alpha5$ initial conditions. On both diagrams three
histograms are plotted: the solid, dashed and dotted lines correspond
to 60\%, 30\% and 15\% star-formation efficiency (the amount of gas
converted into stars) respectively. The dot-dashed line represents the
Salpeter IMF. Masses are shown in M$_\odot$. The bottom panels depict
the cumulative mass distributions from both groups of simulations, on
the left for 15\% efficiency, 60\% efficiency on the right. The
truncated dashed line marks the mass at which the maximum difference
between the two distributions is found. The $\alpha3$ mass
distribution is displayed with diamonds, the $\alpha5$ one with
squares.}  \centerline{\epsfig{file=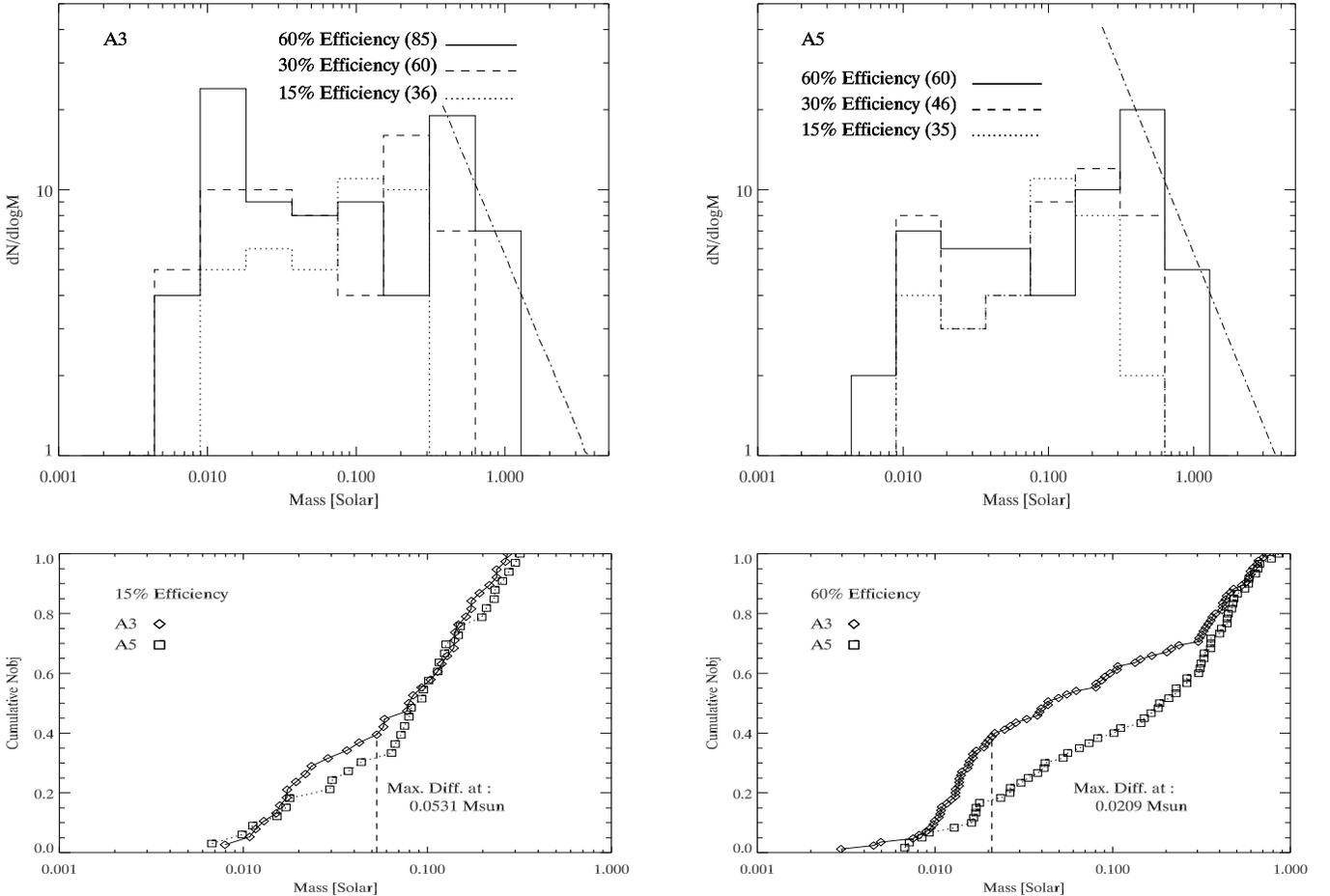,height=13cm}}
\end{center}
\end{figure*}

In Figure~3 [upper panels] we plot the mass function derived by
combining the results from all the simulations of a given $\alpha$. On
the left, the $\alpha3$ case is shown, and the $\alpha5$ mass function
on the right. It can be readily seen that, as mentioned before, the
total number of objects formed is lower in the $\alpha5$ case. In
addition, it is apparent that the fraction of brown dwarfs is also
lower in the $\alpha5$ case, for efficiencies higher than 15\%:
$\approx 30\%$ of all the objects formed in the $\alpha5$ simulations
are brown dwarfs; in contrast, the $\alpha3$ form almost equal numbers
of stars and brown dwarfs. The difference arises most evidently in the
0.01-0.02 M$_\odot$ bin, in which as many as $\approx 20$ brown dwarfs
are located at 60\% efficiency in the $\alpha3$ runs, for only
$\approx 8$ sub-stellar objects in the same bin in the $\alpha5$
simulations. To first order, the transition to the sub-stellar regime
is characterised by a flat slope, though the precise trend is
different depending on the initial value of $\alpha$. For the
$\alpha5$ group, after a turn over at about 0.4 M$_\odot$, the mass
function decreases (in logM space) and has a cutoff at about 0.01
M$_\odot$. The $\alpha3$ mass function displays a similar turn-over
point but later declines in the range 0.02-0.4 M$_\odot$, just to rise
again in the 0.01-0.02 M$_\odot$ bin. A cutoff at $\approx 0.01$
M$_\odot$ is also seen in this case.

The slope at the high mass end ( $ > 0.4 $~M$_\odot$) can be compared
with a Salpeter power-law. The comparison highlights that stars with
masses greater than 1~M$_\odot$ are underproduced, i.e. the high-mass
end does not resemble a Salpeter IMF. This, however, should not be
surprising, since all the clouds have an initial mass of 5 M$_\odot$,
i.e. do not follow any distribution of masses as one would expect in a
real molecular cloud. Delgado-Donate, Clarke \& Bate (2003) showed
that the slope of the IMF before the turn-over is likely to be
dominated and hence follow the slope of the core/cloud mass
function. BBB do find a slope above 0.5 M$_\odot$ which resembles the
Salpeter value, for their simulation of a 50 M$_\odot$ cloud, in which
three entirely independent cores are formed, each with a different
mass. Therefore, our mass functions (MFs) must be understood not so
much as IMFs, but in essence as the building blocks of such IMF. An
IMF could be constructed from our MFs by simply convolving them with a
cloud mass function. For this purpose, our mass functions at the high
mass end may be better described by an approximate $\gamma = 0$ narrow
power law in logM space, followed by a quick drop at about 1
M$_\odot$. This characteristic mass of the upper cutoff depends on the
initial cloud mass, the number of objects formed and the efficiency
assumed.

In order to build an IMF by means of a convolution of the MFs with a
core/cloud mass function we need to make two assumptions, because our
MFs are not scale-free. First, some sort of correlation should exist
between the initial Jeans number and the initial mass of the cloud so
that more massive clouds form more stars. Second, the only mass scale
to influence the pattern of mass acquisition at the high mass end
within each cloud should be the Jeans mass at the opacity limit for
fragmentation and not the initial cloud mass (Bate et al. in
prep.). This last assumption is likely to break down at high cloud
masses ( $> 50$ M$_\odot$), as enough independent sub-clumps may form
so as to yield a mass function readily comparable to the IMF
(BBB). For low-mass star-forming regions, however, this should not be
a concern. Thus, under these two assumptions, we may consider the high
mass end of the mass function per given core mass as being roughly
scale-free and thus amenable to be convolved with a cloud mass
function. If the slope of the cloud mass function is close to
Salpeter's, the result of the convolution will be an IMF that at the
high mass end will resemble closely the observed IMF (see
Delgado-Donate, Clarke \& Bate 2003).

The bottom panels of Figure~3 show in each diagram the cumulative mass
functions for both initial conditions (at 15\% efficiency on the left
and 60\% efficiency on the right). From the cumulative distributions
it is possible to perform a Kolmogorov-Smirnov (KS) test and thus
calculate the probability $P_{\rm KS}$ of both mass functions being
drawn from the same distribution. $P_{\rm KS}$ turns out to be smaller
than 0.05 for efficiencies greater than 15\%. Thus, at the $2\sigma$
confidence level, the two mass functions appear to be
different. However, the mass at which the maximum difference between
the two cumulative distributions occurs (and from which the KS
probability is calculated) are $\approx 0.05$ M$_\odot$ and $\approx
0.02$ M$_\odot$ for the 15\% and 60\% efficiency cases
respectively. [Although large differences are also found in the 60\%
case at larger masses, this only reflects a property of cumulative
distributions: the offset between the two distributions at very low
masses is simply carried forward to larger masses]. Therefore, it is
the mass distribution in the sub-stellar regime that is responsible
for $P_{\rm KS}$ being 0.05. We can conclude then that the stellar
mass function is rather insensitive to the different initial
conditions imposed, whereas the {\it sub-stellar mass function does
depend on the initial slope of the turbulent power spectrum}.

Why does the $\alpha3$ initial condition result in the formation of a
higher fraction of brown dwarfs? The fact that the $\alpha3$
simulations are characterised by having more powerful short wavelength
($\lambda < 1000$ AU) turbulent modes than the $\alpha5$ runs ensures
that small-scale structure is more important in the former. 
Therefore, in the neighbourhood of each collapsing core the gas is
highly structured and thus prone to multiple fragmentation. In
addition, due to the lack of support in large scales against collapse
in the $\alpha3$ case, the accretion rates on to the dense cores and,
in particular, on to the circumstellar discs, is very high, thus
making the discs more likely to be gravitationally unstable than in
the $\alpha5$ case. The net result is that in the $\alpha3$
calculations the number of objects in each small-$N$ cluster is larger
(by a factor of $\approx 2$) than in the $\alpha5$ runs. Consequently
a higher fraction of objects are ejected in the $\alpha3$ case.
Low mass components are the prime candidates for being
ejected, and once they become unbound or simply bound at large
separations, their accretion process is effectively brought to a
halt. Therefore, the $\alpha3$ simulations tend to produce a larger
fraction of brown dwarfs than the $\alpha5$ runs.

In summary, star-forming regions (SFR) which start off with a high
degree of substructure at small scales ($< 1000$ AU) must be expected
to form a large fraction of very low mass objects, and hence render
the sub-stellar mass function different to that of looser, less
structured SFRs. And this should be so independently of the physical
mechanism that drives the formation of such substructure. Thus, by
extension, we speculate that the sub-stellar mass function is likely
to be rather sensitive to the environment in which star formation
takes place, and not exclusively to the slope of the initial power
spectrum, as we have shown explicitly here. This possible dependence
of the sub-stellar IMF on initial conditions may have already been
observed in several star-forming regions. In particular, Brice\~no et
al. (2002) and Preibisch, Stanke \& Zinnecker (2003) find that Taurus
and IC~348 respectively, have a deficit of brown dwarfs, their
fraction relative to stars being $\sim$ a factor of 2 lower than in
more massive SFR such as Orion (Muench et al. 2002), the Pleiades
(Jameson et al. 2002) or $\alpha$ Persei (Barrado y Navascu\'es et
al. 2003).

\subsection{Planetary mass {\it free-floaters} and the minimum mass
for fragmentation}

We mentioned previously that the lower {\it cutoff} in both mass functions
is located at $\approx 0.01$ M$_\odot$. That is, the characteristic
mass at which the number of brown dwarfs drops significantly lies at
about 10 M$_{\rm J}$. No brown dwarf is found to have a mass lower than
$\approx 4$ M$_{\rm J}$, and the fraction of brown dwarfs with masses
below 10 M$_{\rm J}$ is smaller than 10\%. The position of the lower
cutoff in the mass functions is also independent of the choice of
efficiency. Only for very short timescales can brown dwarfs be found
to have masses close to the minimum resolvable mass.

All the objects in our simulations start with a mass close to the
Jeans mass at the critical density $\rho_{\rm c}$, or M$_{\rm res}
\approx 1.5$ M$_{\rm J}$. Therefore, the low fraction of
planetary-mass brown dwarfs is not only determined by our choice of
the critical density $\rho_{\rm c}$ (at which the gas becomes
optically thick) but also by some {\it universal} process occurring to
all collapsed fragments: essentially, the accretion rates during the
first years after each protostar forms are very high, high enough so
as to allow the accretion of $\approx 10 \times$ the initial Jeans
unstable mass before dynamical interactions with neighbouring objects
become important. Typical values for the accretion rates in our
simulations during the first 10 yrs after formation are $\sim
10^{-3}-10^{-4}$ M$_\odot$/yr. We have checked that the initial
accretion rates are not artificially enhanced by a large factor by the
inclusion of sink particles in the calculation. An additional
simulation in which sink particles have an accretion radius $10
\times$ smaller was performed. The accretion rates derived are, on
average, only a factor 5 lower, not low enough to prevent the
accretion of $\approx 0.01$ M$_\odot$ in still a very short timescale.

Thus, the detection of a few planetary-mass free-floating objects
(PMOs) in the mass range $1-10$ M$_{\rm J}$ can be readily
accommodated within the context of our models. However, {\it if} a
{\it large} population of PMOs exists, two explanations for its origin
might be suggested. First, they might form via the same mechanisms at
work in our simulations, namely direct fragmentation in filaments or
disc fragmentation induced by gravitational instabilities. But then
the minimum fragment mass should be $\approx 10 \times$ lower than in
the present simulations, i.e. the critical density $\rho_{\rm c}$ should
be $\approx 100 \times$ higher than assumed here. Although this
possibility cannot be ruled out, such a high value of $\rho_{\rm c}$
would not be easy to explain in terms of our current knowledge of the
thermodynamical processes that determine $\rho_{\rm c}$ (but see Boss
2000).

Second, PMOs may be formed in large numbers in quiescent accretion
discs at later times than modelled here (e.g. during the stellar
dominated phase of star formation), when the disc mass is much smaller
than that of the central object. In this situation, even if
gravitational instabilities drive fragmentation in the disc, the
fragments will lack a large reservoir of gas from which to accrete
substantially, and hence their masses will remain close to the initial
fragment mass, i.e. a few M$_{\rm J}$. These two possibilities are not
mutually exclusive.

\section{Weak variations in the properties of multiple stars}

\begin{figure*}
\begin{center}
\caption{Mass functions for the $\alpha3$ simulations on the left, and
the $\alpha5$ ones on the right. The solid line represents the mass
function calculated using all the stars and brown dwarfs formed in
each group of calculations, at 60\% efficiency. The dashed line
histograms only include the inner members (up to a distance of 1000
AU) of multiples systems, while the dot-dashed line histogram
represents the mass distribution of unbound single objects
(i.e. escapers). Mass is shown in M$_\odot$.}
\centerline{\epsfig{file=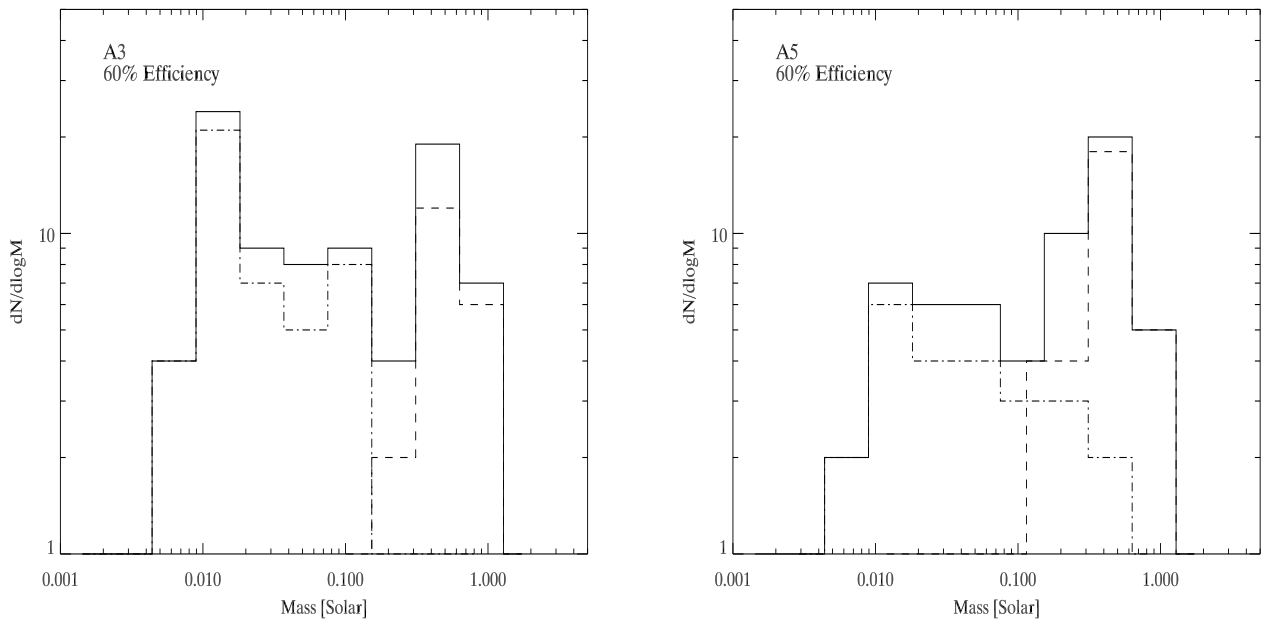,height=9cm}}
\end{center}
\end{figure*}

Delgado-Donate et al. (2003) discussed the properties of the multiple
stars resulting from these simulations. Their analysis referred to the
combined dataset, regardless of which initial conditions had been
applied in each situation. This is justified since the properties of
multiples stars, as we will show presently, display a weak dependence
on the slope of the initial turbulent velocity spectrum.

\subsection{The IMF revisited: companions and singles}

Figure~4 shows the mass function derived from each group of
calculations: on the left the $\alpha3$ case and on the right the
$\alpha5$, at 60\% efficiency. Overplotted we show the mass
distribution of the members of multiple systems up to a separation of
1000 AU (dashed line), and the mass distribution of unbound singles
(i.e. escapers; dot-dashed line). Two features are apparent: first,
the vast majority of the inner members of a multiple system have
high-masses ($0.3 - 1$ M$_\odot$), this being the case independently
of the initial conditions applied. Second, unbound singles (objects
that escaped the cloud after being ejected via a three-body encounter)
populate mostly the $0.01 - 0.1$ M$_\odot$ range. However, a tail of
higher-mass (up to 0.3 M$_\odot$) escapees can be also found in the
$\alpha5$ case. That is, dynamical interactions do not occur so often
in the $\alpha5$ simulations, and therefore some objects have the
chance to accrete to relatively high masses before being ejected from
the system. Nonetheless, it is clear that the pattern of mass
distribution among singles and multiples is very much the same for
both groups of calculations: low-mass singles and high-mass multiple
companions.

Notably, the $\alpha3$ calculations (those in which the number of
stars and brown dwarfs are more similar) are characterised by a mass
function that can be seen as bimodal, the low-mass peak being made up
of single unbound brown dwarfs, and the high-mass peak being populated
by the inner members of multiple systems. This pattern of mass
acquisition was also found in the small-$N$ clusters simulations of
Delgado-Donate, Bate \& Clarke (2003). They performed a large number
of calculations involving a spherical cloud of gas, initially
homogeneous and static, {\it seeded} with 5 accreting point masses
(sink particles) each, and where the processes of competitive
accretion and dynamical interactions could be studied irrespectively
of complications such as star-creation and star-disc
interactions. They found that the bimodal structure of the mass
function per given core could be understood purely as a result of
those two processes: repeated three-body interactions leading to
ejections and binary hardening; and spherical accretion, which was
dubbed competitive because all the stars try to {\it feed} from the
same, finite, gas reservoir. Given the similarity between the
small-$N$ cluster bimodal mass function and the mass function of our
$\alpha3$ simulations, it is tempting to conclude that it is mostly
the interaction between many protostars in a relatively small ($<
10^4$ AU) gas-rich volume that generates this pattern of mass
acquisition, regardless of other intervening processes such as the
initial turbulent flow structure or subsequent star-disc
interactions. In other words, whereas the turbulent flow may determine
the number of interactions between the protostars, and their strength
(by inducing a more or less compact clustering of a lower or higher
number of stars), it is the acting of the dynamical interactions
themselves (i.e. a series of stochastic processes leading to
continuous restructuring of the internal configuration of multiple
systems) in a gas-rich environment what ultimately determines the
properties of multiple stars and the overall shape of the IMF below
the Salpeter range.

\subsection{Orbital parameters}

\begin{figure*}
\begin{center}
\caption{Semi-major axis (in AU) versus primary mass (in M$_\odot$) on
the upper diagrams; eccentricity versus primary mass (in M$_\odot$) on
the bottom ones. On the left, for the $\alpha3$ calculations, on the
right for the $\alpha5$ calculations. The symbol code is as follows:
diamonds represent binaries, triangles triples, squares quadruples,
asterisks quintuples, and crosses higher-order multiples. All the
plots show results at 60\% efficiency. Note: higher-resolution image
available on request to first author.}
\centerline{\epsfig{file=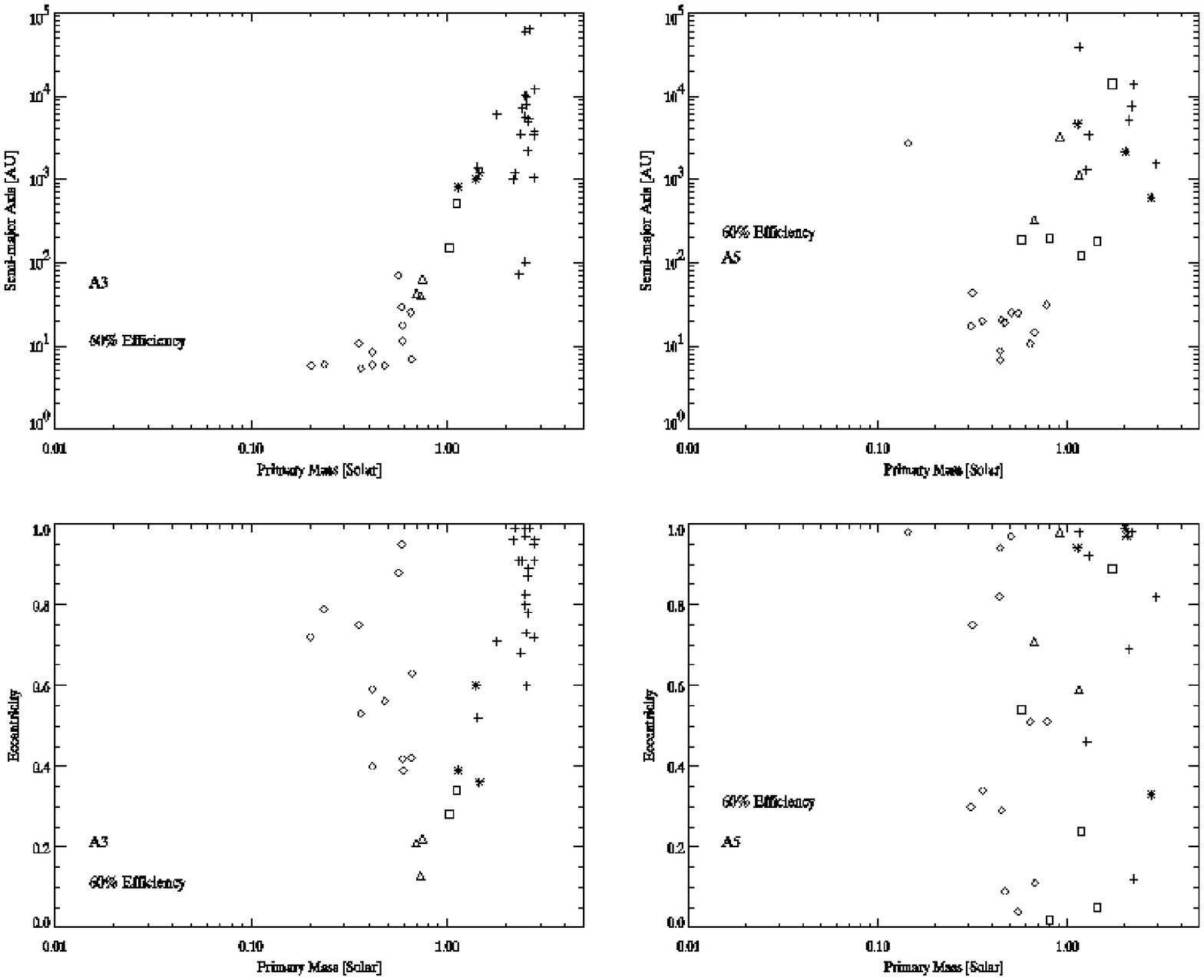,height=15.5cm}}
\end{center}
\end{figure*}

Figure~5 shows the distribution of orbital parameters (semi-major axis
$a$ [top panels] and eccentricity $e$ [bottom panels]) versus primary
mass, for the $\alpha3$ calculations (on the left), and the $\alpha5$
ones, on the right. Figure~5 displays results at 60\% efficiency
only. The symbol code is as follows: binaries are represented by
diamonds, triples by triangles, quadruples by squares, quintuples by
asterisks and higher-order multiples by crosses. Both sets of initial
conditions result in the formation of multiple systems with similar
distributions of $a$ and $e$. The mean separation of binaries is
$\approx 10$ AU, lower than observed for G-stars (Duquennoy \& Mayor
1991; this is due to the lack of wide pure binaries, as described in
Delgado-Donate et al. 2003). Triples, quadruples, etc show an
increasingly larger mean semi-major axis, as it would be expected for
systems that have a hierarchical or nearly hierarchical
configuration. A similar pattern is evident for both groups of
simulations, the most important difference between them being the mean
$a$ of triples, at $\approx 1000$ AU in the $\alpha5$ calculations but
at $\approx 100$ AU in the $\alpha3$ ones.

Binary stars can have a wide range of eccentricities, from 0.05 to
0.95. The $\alpha5$ simulations show a higher incidence of
low-eccentricity binaries, probably due to the lower number of
dynamical interactions that take place in those calculations, relative
to the $\alpha3$ ones. High-order multiples are characterised by very
high values of $e$. These high-order multiples are more frequent in
the $\alpha3$ simulations, simply because many more low mass objects
are formed in this case, and it is the lightest members of an unstable
multiple that are likely to be thrown to wide orbits after ejection.

Overall, although some differences between the distribution of orbital
parameters (as a function of primary mass) for each initial condition
can be appreciated, they are not very significant statistically.

\subsection{Kinematics}

\begin{figure*}
\begin{center}
\caption{Velocity (in km s$^{-1}$) versus primary mass (in M$_\odot$),
at 60\% efficiency. On the left, the $\alpha3$ results; on the right,
the $\alpha5$ results. The symbol code is as in Figure~5, the only
difference being the small tilted crosses, which represent unbound
single objects.}  
\centerline{\epsfig{file=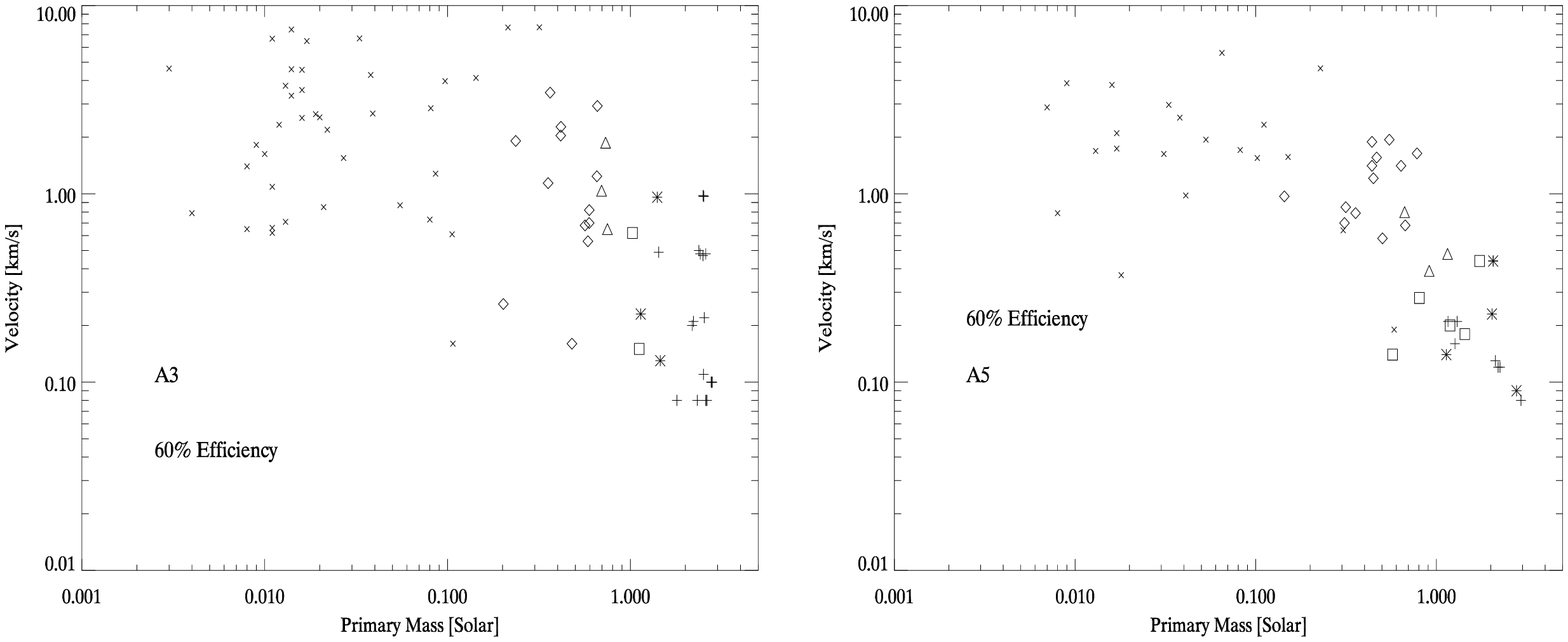,height=8cm}}
\end{center}
\end{figure*}

Figure~6 depicts the distribution of velocities for singles and
multiples, versus primary mass (on the left, for the $\alpha3$ group,
and the $\alpha5$ results shown on the right). The symbol code is as
in Figure~5, with the addition of small tilted crosses, which
represent unbound singles. Both diagrams show similar trends: first,
the highest velocities are attained among singles, but the offset
between the mean velocity of singles and that of multiples (binaries
and triples specifically) is very small. For all practical purposes,
single stars and binaries can be said to define a single kinematical
population. A similar result was also found by BBB in their larger
cluster simulation. However, high-order multiples such as quadruples
or quintuples do have a significantly lower mean velocity. Second,
only a minority of objects (mostly low-mass singles) attain speeds
close to $\approx 10~ {\rm km~s}^{-1}$. The $\alpha3$ case is more
prolific in high-speed escapers. This should be expected, as the
small-$N$ clusters formed in the $\alpha3$ calculations are denser
and, consequently, encounters at shorter distances are more likely.

It must be noted that the number of escapers and their ejection speed
may be sensitive to the softening radius applied in this simulations
($\approx 4$~AU). Following Armitage \& Clarke (1997), we find that
the typical ejection velocity of a very-low-mass star/brown dwarf
which suffers an encounter at 4 AU with a stellar binary is $\approx
10$~km s$^{-1}$. (To calculate this number we have taken the median
mass of escapers and binaries formed in our calculations [0.02
M$_\odot$ and 0.4 M$_\odot$ respectively], and assumed a twin
binary). This 10~km s$^{-1}$ value is substantially higher than the
average velocity of escapers in our simulations, indicating that
encounters at distances of $\approx 4$~AU cannot be frequent. But
particularly in the $\alpha3$ case, there are some escapers with
velocities close to 10~km s$^{-1}$. Therefore, in this case, we might
be missing some encounters with higher velocities. That is not the
case for the $\alpha5$ simulations.

Two conclusions can be drawn from the two trends mentioned above:
first, the majority of low-mass stars and brown dwarfs ever formed in
a high-mass SFR like Orion are expected to remain bound: thus, young
open clusters should still contain most of their initial brown dwarf
population, albeit in the outermost regions due to internal mass
segregation. We see a trend, however, for dense SFRs (our $\alpha3$
simulations) to eject more objects with high velocities than loose
SFRs. Second, the role of binary-binary interactions (leading to the
ejection of binaries) is determinant in mixing the kinematic
properties of single stars and multiples. These interactions could not
take place in the simulations of Delgado-Donate, Bate \& Clarke (2003)
and Sterzik \& Durisen (1998), which formed only one binary per
cluster, and therefore found that the offset between the velocities of
singles and binaries was much larger. Thus, it is only the systems
with highest $N$ (quadruples, quintuples, etc...) that remain
kinematically distinct. These differences are unlikely to be detected
in an actual SFR, since clouds as those we modelled here also move
with respect to each other with a velocity of a few ${\rm
km~s}^{-1}$. But it might be possible that, in loose low-mass
associations such as Taurus, some fraction of the single and {\it
pure} binary population has escaped to the field. This might help to
explain why the binary fraction in Taurus is enhanced by a factor of 2
compared to solar-type stars on the main sequence (Ghez, Neugebauer \&
Matthews 1993; Simon et al. 1995; K\"ohler \& Leinert 1998), {\it and}
why the companion frequency (the number of companions per stellar
system) is the highest of all young SFRs (Duch\^ene 1999).

\section{Conclusions}

We have undertaken a series of hydrodynamical simulations of multiple
star formation in small molecular clouds. Our approach of modelling 10
independent small clouds of 5 M$_\odot$ each instead of just one cloud
of 50 M$_\odot$ has allowed us to explore different initial
conditions. In this paper we have discussed the effect that different
slopes of the power law spectrum of the initial turbulent velocity
field has on the properties of the resulting stars and brown
dwarfs. Two slopes have been applied, $\alpha = -3$ and $\alpha = -5$,
and the other initial parameters (total mass, cloud radius, number of
Jeans masses, initial turbulent kinetic energy) have been kept
constant. Particular emphasis has been given to the analysis of the
mass function of the stars and brown dwarfs formed in these
simulations, and its possible variation with initial conditions. It is
worth mentioning that the IMFs analysed in this paper are derived {\it
directly} from the masses of stars actually formed in numerical
calculations, as a result of the interplay between turbulence,
self-gravity, competitive accretion between protostars and dynamical
interactions in unstable multiple systems. In our models, stars and
brown dwarfs start off with masses close to the opacity limit for
fragmentation (a few M$_{\rm J}$) and subsequently grow in mass by
accretion. Our approach is different to that of Padoan \& Nordlund
(2002) who, although they started with isothermal gas and similar
inputs of turbulent energy, did not include self-gravity in their
simulations nor followed fragmentation down to the opacity limit. They
derived an IMF by applying some relations concerning the Jeans mass to
the density probability distribution function of compressible
turbulence. No actual self-gravitating object (i.e.`star') was (or
indeed could be) formed in their calculations.

Our main conclusions are:
\begin{itemize}
\item The fraction of brown dwarfs (out of the total number of stellar
and sub-stellar objects) formed in our calculations is sensitive to
the initial slope of the turbulent power spectrum. The brown dwarf
fractions produced by the 2 sets of simulations are statistically
different at the $2\sigma$ level. The origin of this difference can be
explained in terms of the degree of substructure that the different
initial conditions are able to generate. In the $\alpha3$ case, the
amount of kinetic energy stored in short-wavelength ($\lambda < 1000$
AU) turbulent modes is higher than in the $\alpha5$ case, and
consequently the dense cores in which the cloud fragments remain
highly structured even after the decay of turbulence. This results in
the formation of more objects per dense core in a more compact
configuration, leading to a higher incidence of ejections of low-mass
members than in the $\alpha5$ case. Therefore, we speculate that the
shape of the sub-stellar mass function is likely to be sensitive to
the degree of substructure present in each SFR (observational hints in
this direction have recently been provided by Brice\~no et al. 2002,
Luhman et al. 2003 and Preibisch, Stanke \& Zinnecker 2003),
independently of the physical process responsible of the generation of
such substructure. A KS test applied to the mass functions resulting
from each $\alpha$ also demonstrates that the distribution of masses
at the stellar regime does not show any significant dependence on the
value of $\alpha$. Thus, we conclude that it is the slope of the
sub-stellar IMF, rather than that of the stellar IMF, that is likely
to be affected by star-formation environmental conditions.
\item We find that few brown dwarfs with masses less than $\approx
0.01$ M$_\odot$ are formed in our simulations. Only 10\% of all brown
dwarfs have masses in the range $1-10$ M$_{\rm J}$, despite the
minimum fragment mass being $10 \times$ lower. This result is a
consequence of the high accretion rates ($\sim 10^{-3}-10^{-4}$
M$_\odot$/yr) characteristic of the very first years of a protostar's
life, which result in the mass of most of the objects increasing by a
factor of 10 in a very short timescale. This timescale is typically
shorter than $\sim 100$ yr and therefore the probability that
dynamical interactions act to eject the object during that time is
almost negligible. Therefore we conclude that although the detection
of a few planetary-mass free-floating objects (PMOs) can be
accommodated by our models, if a {\it large} population of PMOs exists,
then other explanations for their origin must be sought. Either the
value of the critical density $\rho_{\rm c}$ that determines the
minimum fragment mass may need to be revised, and/or PMOs may be
formed in large numbers in quiescent discs at later stages than
modelled here -- in a relatively gas-poor environment --, when the
disc mass is much smaller than that of the central object and
therefore a large reservoir of gas is not available for the fragment
to grow in mass by accretion.
\item The pattern of mass acquisition among single and
multiple systems is shown not to depend sensitively on the slope of
the initial turbulent power spectrum. Likewise, the distribution of
orbital parameters (semi-major axis and eccentricity) of the multiples
is only weakly dependent on initial conditions.
\item Singles and binaries constitute a kinematically homogeneous
population (mean velocity dispersion $\sim$ few km s$^{-1}$). The
offset between the mean velocity dispersion of each group is
substantially smaller than found in the simulations of Delgado-Donate,
Clarke and Bate (2003), and arises from the mutual interactions
between binary systems. High-order multiples ($N > 4$) attain
significantly lower velocities ($\sim 0.1$ km s$^{-1}$), and thus
might remain closer to the densest cores of a SFR than higher-speed
members. Only a minority of objects (mostly low-mass singles) attain
speeds close to $\approx 10 {\rm km~s}^{-1}$. The $\alpha3$ case is
more prolific in these high-speed escapers, indicating that as
expected, dense SFRs eject more objects with high velocities than
loose SFRs.
\end{itemize}

\section*{Acknowledgments}

EJDD is grateful to the EU Research Training Network {\it Young
Stellar Clusters} for support. CJC gratefully acknowledges support
from the Leverhulme trust in the form of a Philip Leverhulme Prize. We
thank the referee, Gilles Chabrier, for useful comments which helped
us to improve the paper. The computations reported here were performed
using the U.K. Astrophysical Fluids Facility (UKAFF).

\label{lastpage}

\end{document}